\documentclass[12pt]{article}
\usepackage{epsfig}
\usepackage{color}
\usepackage{amssymb,amsmath}
\usepackage{bm}
\usepackage{graphicx}

\newcommand{\bea}{\begin{eqnarray}}
\newcommand{\eea}{\end{eqnarray}}

 \makeatletter
\def\alt{\mathrel{\mathpalette\gl@align<}}
\def\agt{\mathrel{\mathpalette\gl@align>}}
\def\gl@align#1#2{\lower.6ex\vbox{\baselineskip\z@skip\lineskip\z@
\ialign{$\m@th#1\hfil##\hfil$\crcr#2\crcr\sim\crcr}}} \makeatother

\begin{document}
\begin{flushright}
KEK-TH-1327

OU-HET 762/2012
\end{flushright}
\vspace*{1.0cm}

\begin{center}
\baselineskip 20pt 
{\Large\bf 
TeV Scale $B-L$ model   with \\ a flat Higgs potential 
at the Planck scale

\vspace{.5cm}
- in view of the hierarchy problem -
}
\vspace{1cm}

{\large 
Satoshi Iso$^{a,b,}$\footnote{satoshi.iso@kek.jp}
and 
Yuta Orikasa$^{c,}$\footnote{orikasa@het.phys.sci.osaka-u.ac.jp}
} \vspace{1.0cm}

{\baselineskip 20pt \it
$^{a}$  Theory Center, 
High Energy Accelerator Research Organization (KEK)  \\
$^{b}$
The Graduate University for Advanced Studies (SOKENDAI), \\
1-1 Oho, Tsukuba, Ibaraki 305-0801, Japan \\
and \\
$^{c}$Department of Physics, 
Osaka University , \\ Toyonaka,
Osaka 560-0043, Japan
}

\vspace{1.0cm} {\bf Abstract} \\
\end{center}
The recent discovery of the Higgs-like particle at around 126 GeV 
has given us a big hint  towards the origin of the Higgs potential.
Especially the  running quartic coupling
 vanishes near the Planck scale, which
 indicates a possible link between the physics in the electroweak and the Planck scales.
Motivated by this and the hierarchy problem, 
we investigate a possibility that the Higgs has a flat potential at the Planck scale.
In particular, we study the RG analysis of the 
$B-L$  extension of the standard model \cite{IOO} with a classical conformality.
The $B-L$ symmetry is radiatively broken at the TeV scale via the
Coleman-Weinberg mechanism.
The electroweak symmetry breaking  is triggered by a radiatively generated
scalar mixing so that its scale $246$ GeV is dynamically related with the 
$B-L$ breaking scale at TeV.
The Higgs boson mass is given at the border of the stability bound,
which is lowered by a few GeV from the SM 
by the effect of  the $B-L$ gauge interaction.

\thispagestyle{empty}

\newpage

\addtocounter{page}{-1}
\setcounter{footnote}{0}
\baselineskip 18pt
\section{Introduction}
The dynamics of the electroweak symmetry breaking (EWSB) and the origin of the Higgs potential
are the most important issues in  the standard model (SM). 
The ATLAS and CMS groups have announced a discovery 
of the Higgs-like particle at around 126 GeV \cite{ATLAS,CMS}. This value of 126 GeV is quite
 suggestive to the physics at very high energy 
since it is close to the border of the vacuum stability bound up to the Planck scale. 
In the SM, the Higgs mass is determined by the quartic coupling $\lambda_H$ of the 
Higgs field. For a relatively light Higgs boson, the $\beta$ function of $\lambda_H$ becomes negative and
the running coupling $\lambda_H$ crosses zero at some high energy scale. This implies an instability 
of the Higgs potential.
The theoretical investigation \cite{Espinosa} 
shows that the stability bound up to the Planck scale requires (see also \cite{topmass} for larger uncertainties 
of the top mass)
\begin{align}
M_h[\mbox{GeV}] > 129.4 + 1.4 \left( \frac{M_t[\mbox{GeV}]-173.1}{0.7}\right)
 -0.5  \left( \frac{\alpha_s(M_Z)-0.1184}{0.0007}\right) \pm 1.0_{th} .
\label{stability-bound}
\end{align}
The observed mass of 126 GeV is close to but lower by a few GeV from the above value
of the stability bound in the SM. If the Higgs mass is lighter than the above bound, new physics must appear
below the Planck scale. But if it is just at the border of the stability bound, 
it may give a big hint to the origin of the Higgs potential at the Planck scale 
\cite{FP1,FP2,FP3,FP4}.

Another important clue to the Higgs potential comes from the hierarchy problem, i.e. the 
stability of the Higgs mass against higher energy scales
such as the GUT or the Planck scale. 
The most natural solution is the low energy supersymmetry, but
the new physics search at the LHC  
has put  stringent constraints on simple model constructions and
a large parameter region of the TeV scale supersymmetric models  has been already excluded.
Of course,  we cannot rule out a possibility to find 
an indication of the low energy supersymmetry in the near future, 
but it will be important to 
reconsider the hierarchy problem from a different point of view.

In this paper, we take an alternative approach to the hierarchy problem following the Bardeen's argument
\cite{Bardeen}.
In section 2, we give an interpretation of his argument in terms of the renormalization group equations (RGE).
If we adopt it, the most natural mechanism to break the electroweak symmetry
is the Coleman-Weinberg (CW) mechanism \cite{CW}. In section 3, we emphasize that the CW mechanism
is another realization of a dimensional transmutation, 
and  stable against higher energy scales.
It is, however, well known that the CW mechanism does not work within the SM because of 
the large top Yukawa coupling. Hence we need to extend the SM.
In section 4, we introduce our model, a classically conformal $B-L$ extension of the SM \cite{IOO}.
The anomaly free global symmetry of the SM, $B-L$ (baryon number minus lepton number), is gauged,
and the right-handed neutrinos and a SM singlet scalar $\Phi$ are introduced. 
This model has a classical conformal invariance  \cite{MaNi}, namely there are no explicit mass terms in the scalar
potential. We furthermore assume, motivated by the 126 GeV Higgs, that the Higgs potential
is flat at the Planck scale.
In section 5, we discuss the dynamics of the model using the RGEs. 
We first study the 
radiative breaking of the $B-L$ gauge symmetry via the CW mechanism.
Then we show that a small negative value of the mixing 
$\lambda_{mix} (H^\dagger H)(\Phi^\dagger \Phi)$
is radiatively generated  by solving the RGEs.
This triggers the EWSB.
We then discuss the predictability of the model. 
We also show that the stability bound of the Higgs potential up to 
the Planck scale is lowered than the SM prediction by a few GeV by the $B-L$ gauge interaction.
Finally we conclude in section 6.

\section{Bardeen's argument on the hierarchy problem} 
We pay a special attention to the almost scale invariance of the SM.
At the classical level, the SM Lagrangian is conformal
invariant except for the Higgs mass term.
Bardeen has argued \cite{Bardeen} that once the classical conformal invariance and its minimal violation
by quantum anomalies are imposed on the SM, it may be free from the quadratic divergences.
Bardeen's argument on the hierarchy problem is interpreted as follows \cite{AokiIso}.
In field theories, we have two kinds of divergences, logarithmic and quadratic divergences.
The logarithmic divergence is operative both in the UV and the IR. In particular, it  controls a
running of coupling constants and is observable. On the other hand, the quadratic divergence can be
always removed by a subtraction. Once subtracted, it no longer appears in observable quantities.
In this sense, it gives a boundary condition of a quantity in the IR theory at the UV energy scale
where the IR theory is connected with a UV completion theory. 
Indeed, the RGE of a Higgs mass term $m^2$  in the SM 
\begin{align}
V(H)= - m^2 H^\dagger H + \lambda_H (H^\dagger H )^2
\end{align}
is approximately given by
\begin{align}
\frac{dm^2}{dt}=\frac{m^2}{16 \pi^2}
\left( 12 \lambda_H + 6 Y_t^2 - \frac{9}{2}g^2
-\frac{3}{2}g_Y^2 
\right).
\end{align}
$Y_t$ is the top Yukawa coupling and $g, g_Y$ are $SU(2)_L, U(1)_Y$ gauge couplings. 
The quadratic divergence is adjusted by
 a boundary condition either at the IR or UV scale. Once 
the initial condition of the RGE is given at the UV scale, it is no longer operative in the IR.
The RGE shows that the mass term $m^2$ is multiplicatively renormalized.
If it is zero at a UV scale $M_{UV}$, it continues to be zero at lower energy scales.
In this sense, the quadratic divergence is not the issue 
in the IR effective theory, but the issue of the UV completion theory.

The multiplicative renormalizability of the mass term is violated by
a presence of a mixing with another scalar field $\Phi$
\begin{align}
V_{mix}(H,\Phi) = \lambda_{mix} (H^\dagger H)(\Phi^\dagger \Phi).
\end{align}
Then the RGE is modified as
\begin{align}
\frac{dm^2}{dt}=\frac{m^2}{16 \pi^2}
\left( 12 \lambda_H + 6 Y_t^2 - \frac{9}{2}g^2
-\frac{3}{2}g_Y^2 
\right)  + \frac{M^2}{8\pi^2} \lambda_{mix}
\label{RGEm}
\end{align}
where $M$ is the mass of the $\Phi$ field. The last term comes from the 
logarithmic divergence 
\begin{align}
\delta m^2 \sim  \frac{\lambda_{mix} M^2}{16\pi^2} \log(M^2/m^2)
\label{RGEm2}
\end{align}
due to the  loop diagram of the scalar particle $\Phi$.
Therefore, the hierarchy problem, namely the stability of the EWSB scale, is 
caused by such a mixing of relevant operators (mass terms) with  hierarchical energy scales
$m \ll M$. 

From the above considerations,
we can divide  the hierarchy problem  into the following two different issues;
\begin{itemize}
\item Boundary condition of   dimensionful parameters (such as $m^2$) at $M_{UV}$
\item Mixing of relevant operators as in (\ref{RGEm})
\end{itemize}
The first is related to the quadratic divergences while the second to the logarithmic divergences.

Supersymmetry is  most favored  in solving the hierarchy problem.
If its breaking scale is not so high, it can solve both issues of the hierarchy problem.
It is beautiful, but the recent experiments have put severe constraints on 
the model constructions with the low energy supersymmetry.
But we do not need to solve both issues simultaneously.
Quadratic divergences are subtracted at the UV cut-off scale
as a boundary condition. The justification is necessary in
the UV completion theory. 
On the contrary, 
in order to avoid the  operator mixings with high energy scales, 
we need to impose an absence of
 intermediate scales  between TeV and Planck
scales\footnote{This requires that the grand unification, if exists, occurs  at the Planck scale. 
It is interesting to construct  phenomenologically viable models 
in string theory that both of the supersymmetries
and the grand unification are broken at the Planck scale.}.
This is emphasized in the Bardeen's argument \cite{Bardeen}.
Then the  Planck scale physics is directly connected with the electroweak physics. 
Such a view has been  emphasized  also by Shaposhnikov \cite{shapo,shapo2}.
A natural boundary condition of the mass term 
at the UV cut-off scale, e.g. $M_{Pl}$, is
\begin{align}
 m^2(M_{Pl}) =0.
\label{boundary-condition}
\end{align}
This is the condition of the classical conformality.
The condition (\ref{boundary-condition}) must be justified in the UV completion theory.
From the low energy effective theory point of view, it is just imposed
as a boundary condition.

\section{Stability of the Coleman-Weinberg mechanism}
If there are no intermediate scales, mass parameters are multiplicatively renormalized.
Then, if we set the dimensionful mass parameters
 zero at the UV scale, they continue to be absent in the low energy scale. 
Such a model is called a {\it classically conformal} model  \cite{MaNi}.
Conformal invariance 
is broken  by a logarithmic running of the coupling constants, but no explicit mass terms 
arise by radiative corrections.
Hence, the EWSB must be realized not by the negative mass squared
term of the Higgs doublet but by the radiative breaking
such as the Coleman-Weinberg (CW) mechanism \cite{CW}. 
In this section we see the stability of the symmetry breaking scale in the CW mechanism
against higher energy scales.

In the SM, we have two typical mass scales, QCD  and the electroweak
scales. Let us compare the emergence of a low energy scales in the CW mechanism and QCD.
QCD scale $\Lambda_{QCD}$ is  dynamically generated at a low energy  where
the running coupling constant diverges. It is given as a function of
the coupling $\alpha_s(M_{UV})$ at a UV scale $M_{UV}$ by
\begin{eqnarray}
\Lambda_{QCD} =M_{UV} \exp \left(-\frac{ 2\pi} {b_0 \alpha_s(M_{UV})}\right).
\label{QCD}
\end{eqnarray}
 $b_0$ is the coefficient of the $\beta$ function $\beta=d\alpha_s/dt=-(b_0/2\pi) \alpha_s^2$.
Since the $\beta$ function is proportional to $\hbar$, the small QCD scale $\sim M_{UV} \exp(-c/\hbar)$
is nonperturbatively generated, and stable against radiative corrections of higher energy scales.

Similarly, if the EWSB is realized by the CW mechanism,
its breaking scale  emerges radiatively from the coupling constant at a UV scale.
In comparison to the dimensional transmutation in QCD, the symmetry breaking scale $M_{CW}$
emerges  near the scale where the running coupling constant crosses zero.
In order to realize the zero-crossing of the running coupling constant, 
the $\beta$ function must take a positive value  $\beta>0$ near the breaking scale $M_{CW}$.
Let's make an approximation that $\beta=b>0$ is constant for simplicity.
See, e.g. eq.(\ref{RGElammix-sim}). Then the running coupling constant is approximately given by
\begin{align}
\lambda(t)=b(t-t_0)=b(t-t_{UV})+\lambda_{UV},
\end{align}
where we have introduced the boundary condition $\lambda(t_{UV})=\lambda_{UV}$.
The running coupling $\lambda(t)$ vanishes at $t=t_0 =t_{UV}-\lambda_{UV}/b$.
The renormalized effective potential of the scalar field $\phi$ with a quartic self-coupling $\lambda$
is given by 
\begin{eqnarray}
V(\phi)=\frac{1}{4} \lambda(t)  \phi^4 =\frac{bM^4}{4}(t-t_0) e^{4t}
\end{eqnarray} 
where $t=\log [\phi/M]$ and $M$ is the renormalization point. 
(We have neglected the anomalous dimension of the field for simplicity.) 
The potential has a minimum at $t=t_0-1/4$.
Hence, the  breaking scale 
 $M_{CW}=\langle \phi \rangle$ is given by
\begin{eqnarray}
 M_{CW} = M_{UV} \exp (-\frac{\lambda_{UV}}{b} -\frac{1}{4}). 
\label{CWscale}
\end{eqnarray}
The emergence of the scale $M_{CW}$ is similar to the dimensional transmutation
(\ref{QCD})  in QCD.
The exponent of the rhs in eq.(\ref{CWscale}) shows a balance between the contribution
to the effective potential from  the tree-level coupling 
$\lambda_{UV}$ and the loop contribution proportional to $b$ (and $\hbar$).  
Such a balance is necessary for the CW mechanism to occur. 
In particular, as emphasized in \cite{CW}, the CW mechanism does not occur in a scalar QED
without the gauge interaction.
A small value of $b$  can
generate  a small energy scale $M_{CW}$ from a very high energy scale $M_{UV}$.
In this sense, the CW mechanism is similar to the dimensional transmutation in QCD and stable
against the higher energy scales.
It is the reason why the CW mechanism can be an alternative solution to the gauge hierarchy problem.

\section{Classically conformal $B-L$ model}
In the SM, the dominant contribution to the $\beta$ function of the Higgs quartic coupling
comes from the gauge couplings, top Yukawa
coupling  and the quartic coupling itself,
\begin{align}
\beta_H
 = \frac{1}{16 \pi^2}
\left( 24 \lambda_H^2 - 6 Y_t^4 + \frac{9}{8}g^4 + 
\frac{3}{8}g_Y^4  \right).
\label{betaSM}
\end{align}
In order to realize the CW mechanism, the $\beta$ function must take a positive value.
It is, however, well-recognized that the large top-Yukawa coupling
$Y_t$ makes it negative and the CW mechanism does not work in the SM.
Hence, in order to break the EW symmetry radiatively, 
we need to extend the SM so that the CW mechanism works with 
phenomenologically viable parameters.

The idea to utilize the CW mechanism to solve the hierarchy problem was first
modelled by Meissner and Nicolai \cite{MaNi}.
They proposed an extension of the SM with 
the classically conformal invariance
(see also 
\cite{Dias,Hempfling,Wu,Foot,Holthausen,AlexanderNunneley,Ishiwata,Lee}). 
In addition to the SM particles, right-handed neutrinos and a SM singlet scalar $\Phi$ are 
introduced.

In previous papers \cite{IOO}, inspired by the above work \cite{MaNi},
we have proposed a minimal 
phenomenologically viable model that the electroweak symmetry 
can be radiatively broken.
It is the minimal $B-L$ model \cite{B-L,B-L2,Basso}.
The model is similar to the one proposed by \cite{MaNi}, but the difference
is whether the $B-L$ symmetry is gauged or not. 
We showed that the gauging of $B-L$ symmetry
plays an important role to achieve the radiative $B-L$ symmetry breaking.
It is also phenomenologically favorable. 

\subsection{The model}
A classically conformal $B-L$ extension of the SM is based on 
the gauge group $SU(3)_c \times SU(2)_L \times U(1)_Y \times U(1)_{B-L}$.
The particle contents are listed (except for the gauge bosons) in Table 1.
\begin{table}[t]
\begin{center}
\begin{tabular}{c|ccc|c}
            & SU(3)$_c$ & SU(2)$_L$ & U(1)$_Y$ & U(1)$_{B-L}$  \\
\hline
$ q_L^i $    & {\bf 3}   & {\bf 2}& $+1/6$ & $+1/3$  \\ 
$ u_R^i $    & {\bf 3} & {\bf 1}& $+2/3$ & $+1/3$  \\ 
$ d_R^i $    & {\bf 3} & {\bf 1}& $-1/3$ & $+1/3$  \\ 
\hline
$ \ell^i_L$    & {\bf 1} & {\bf 2}& $-1/2$ & $-1$  \\ 
$ \nu_R^i$   & {\bf 1} & {\bf 1}& $ 0$   & $-1$  \\ 
$ e_R^i  $   & {\bf 1} & {\bf 1}& $-1$   & $-1$  \\ 
\hline 
$ H$         & {\bf 1} & {\bf 2}& $-1/2$  &  $ 0$  \\ 
$ \Phi$      & {\bf 1} & {\bf 1}& $  0$  &  $+2$  \\ 
\end{tabular}
\end{center}
\caption{
Particle contents of minimal $B-L$ model (except for the gauge bosons).
In addition to the SM particles, 
the right-handed neutrino $\nu_R^i$ 
($i=1,2,3$ denotes the generation index) 
and a complex scalar $\Phi$ are introduced. 
}
\end{table}
In addition to the SM particles, the model consists of the
$B-L$ $U(1)$ gauge field, a SM singlet scalar $\Phi$ and right handed neutrinos $\nu^i_R$.
The three generations of right-handed neutrinos ($\nu_R^i$)
are necessary to make the model free from all 
the gauge and gravitational anomalies. 
The Lagrangian relevant for the seesaw mechanism is given by
\bea 
 {\cal L} \supset -Y_D^{ij} \overline{\nu_R^i} H^\dagger \ell_L^j  
- \frac{1}{2} Y_N^i \Phi \overline{\nu_R^{i c}} \nu_R^i 
+{\rm h.c.},  
\label{Yukawa}
\eea
where the first term gives the Dirac neutrino mass term 
after the EWSB, 
while the right-handed neutrino Majorana mass term 
is generated through the second term associated with 
the $B-L$  symmetry breaking. 
Without loss of generality, we here work on the basis
where the second term is diagonalized and 
$Y_N^i$ is real and positive. 

The gauge couplings are introduced in the covariant derivative\footnote{Since there are multiple matter fields
coupled differently with the two gauge fields in the present model, we need to introduce  $g_{mix}$
as an independent parameter. See \cite{gaugemixing} for a general discussion.} 
\begin{eqnarray}
D_\mu \phi= \partial_\mu \phi + i[g_Y Q^Y B_\mu + (g_{mix} Q^Y +g_{B-L} Q^{B-L})B_\mu^{B-L}].
\end{eqnarray}
Here $B_\mu$ and $B_\mu^{B-L}$ are gauge bosons of $U(1)_Y$ and $U(1)_{B-L}$, and
 $Q^Y$, $Q^{B-L}$ are their charge operators. 
These $U(1)$  gauge bosons get mixed through loop corrections. 
The scale at which the U(1) gauge mixing becomes zero is introduced as a new parameter.
The magnitude of the running coupling $g_{mix}$ is mostly determined by
 other gauge couplings $g_Y$ and $g_{B-L}$.
In the previous analysis of the model in \cite{IOO}, we neglected the effect of the gauge mixing 
between $U(1)_{B-L}$ and $U(1)_Y$. 
Since we are interested in the radiative mixing of the scalars, $\Phi$ and $H$,
the mixing of $U(1)$ gauge bosons plays a very important role in this paper.

\subsection{Higgs potential}

Under the hypothesis of the classically conformal invariance, the  scalar 
potential is  given by
\bea 
 V(\Phi, H) = \lambda_H (H^\dagger H)^2 + \lambda_{\Phi} (\Phi^\dagger \Phi)^2 
   + \lambda_{mix} (\Phi^\dagger \Phi) (H^\dagger H).  
\label{potential}
\eea
In \cite{IOO},  we chose these 3 quartic couplings 
by hand so that the $B-L$ and EW symmetries are spontaneously broken at $TeV$ and EW scales.
Especially  it was necessary to take the  mixing  $\lambda_{mix}$ 
 to be a small negative value $\sim (- 10^{-3})$, which seems quite artificial. 
In this paper, we show that such a small negative scalar mixing can be radiatively generated
if we assume that the Higgs has a flat potential at a UV scale (e.g. Planck scale).

Now we explain the most important assumption in the paper.
Motivated by the light Higgs boson mass around 126 GeV, 
we impose a simple assumption that \\
 \centerline{ {\it The Higgs has a flat potential  at a UV scale, $M_{UV}$}. }
The vanishing  quartic coupling $\lambda_H$ at a high energy scale
is suggested  by the experimental indication of the light Higgs boson mass around 126 GeV.
Namely, the  running coupling $\lambda_H (t)$ crosses zero at a UV scale,
and the vacuum becomes unstable.
In  \cite{Espinosa}, a detailed investigation  is presented, 
and the Higgs potential is shown to develop an instability around $10^{11}$ GeV
for the Higgs mass $124-126$ GeV. However, because of the very slow running of the Higgs quartic coupling
at a higher energy scale, the instability scale  is very sensitive to  
theoretical and experimental uncertainties, and the
stability up to the Planck scale cannot be excluded
(see recent papers \cite{topmass}). 
Here we take the light Higgs boson mass as an indication of a vanishing quartic coupling at the UV scale
$M_{UV}$. 

In addition to it, we further assume that the scalar mixing 
$ \lambda_{mix} H^\dagger H |\Phi|^2$  vanishes at $M_{UV}$. Then the 
 potential $V(\Phi, H)$ is completely flat
into the direction of $H$ and   becomes
\bea 
 V(\Phi, H)|_{UV} =  \lambda_{\Phi} (\Phi^\dagger \Phi)^2 
\label{potentialPL}
\eea
at the UV scale $M_{UV}$. We will show in the following section that
radiative corrections generate a small negative
value of the scalar mixing  $\lambda_{mix} \sim -10^{-3}$  at a lower energy scale. 
If the  $B-L$ symmetry is broken, 
its VEV $\langle \Phi \rangle$ triggers the 
EWSB. 
Hence the scales of $B-L$ breaking and the EWSB are related.
The square root of the mixing $|\lambda_{mix}|$
gives a ratio between these two scales.

Because of  the classically conformal invariance and the assumption that the Higgs has a flat potential
at the UV scale, the model is characterized by a very few parameters.
Besides the SM couplings and the Yukawa couplings of $\nu_R$, 
the model has  three additional parameters
\begin{enumerate}
\item $B-L$ gauge coupling ($g_{B-L}$)
\item SM singlet quartic coupling ($\lambda_\Phi$)
\item Energy scale at which $g_{mix}$ vanishes.
\end{enumerate}
As stated above, the magnitude of the gauge mixing is almost determined by the magnitudes of
other gauge couplings
and the scale at which $g_{mix}$ vanishes is not very important in determining the dynamics.
In this sense, there are only two parameters that are important in the dynamics of the model.
One of  them determines the scale of the EWSB.
Hence the model is essentially described by only one parameter and has a high
predictability (or excludability). 
\section{RGE analysis of the model} 
In this section, we  look at the  behaviors of  the  RGEs, 
and discuss how the symmetry breakings occur.
The RGEs are given in the appendix and they can be easily solved numerically.

The classical conformality forbids the explicit breaking of the conformal invariance
by dimensionful parameters. So no scalar mass terms
are allowed at the UV cut-off scale.
If it is absent at the UV boundary, it no longer appears at a lower energy scale
as shown in  eq.(\ref{RGEm}). 
We further impose the flat potential hypothesis.
Then the scalar potential is given by eq.(\ref{potential})
with a boundary condition 
\begin{align}
\lambda_{H}(M_{UV}) = \lambda_{mix}(M_{UV}) =0.
\end{align}
$V(\Phi,H)$ is invariant under the shift of $H$ at the cut off scale $M_{UV}$.
But the shift symmetry is violated by the gauge coupling
of the Higgs field or Yukawa couplings.
They generate the potential of the Higgs field at a lower energy scale through radiative corrections.  
The only parameter in the scalar potential at the Planck scale is the quartic coupling  $\lambda_\Phi$ of the 
SM singlet scalar field, which determines
the $B-L$ breaking scale $M_{B-L}$ via the CW mechanism. 
\subsection{$B-L$ symmetry breaking}
Let us first look at the behavior of the quartic coupling of the $\Phi$ field. 
The RGE is given by eq.(\ref{RGElamphi}). For appropriate parameters,
the self-coupling $\lambda_\Phi$ is positive at a higher energy scale and crosses
 zero at a lower scale.
The running coupling $\lambda_\Phi(t)$ near the crossing point
is proportional to the (positive-valued) $\beta$ function,
\begin{align}
\alpha_{\lambda,\Phi}\equiv \frac{\lambda_\Phi}{4 \pi} \propto \alpha^2_{B-L} > \frac{1}{96} Tr[\alpha_N^2].
\end{align}
The  inequality is required by the positiveness of the $\beta$ function.
If it is satisfied, $\Phi$ field gets a nonzero vev and the $B-L$ symmetry is spontaneously 
broken. See \cite{IOO} for more details. 
The breaking scale is correlated with the  coupling $\lambda_\Phi(M_{UV})$
at a UV cut-off.
\begin{figure}[t]\begin{center}
\includegraphics[scale=.6]{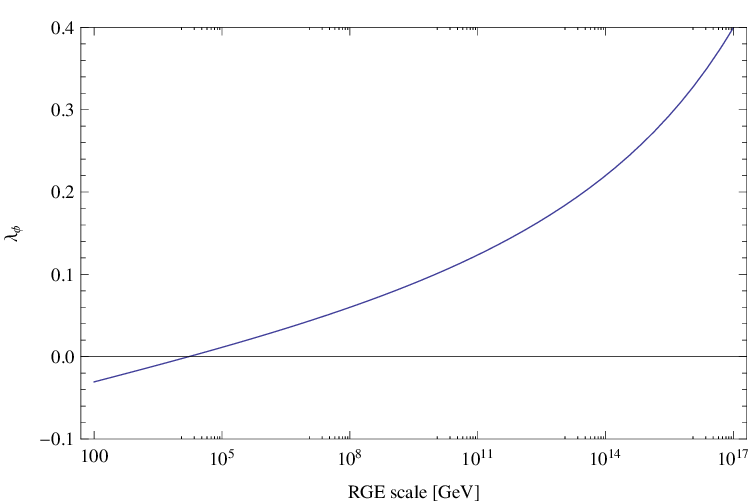}
\label{fig:self}
\caption{
RG evolution of the self-coupling $\lambda_\phi$ of a SM singlet scalar $\phi$.
Since the $\beta$ function is positive, the running coupling crosses zero
at a lower energy scale.
}
\end{center}
\end{figure}
The figure 1 shows a typical behavior of the running coupling $\lambda_\Phi$.  
For a specific choice of $\lambda_{\Phi}$ at the Planck scale,
it crosses zero at a lower energy scale around $M_0 \sim \ 10 \ TeV$.
Then the $B-L$ symmetry is broken at $M_{B-L} \sim M_0 \exp(-1/4)$.  
As shown in \cite{IOO}, the SM singlet scalar has a mass
\begin{align}
m_\phi^2 = \frac{6\pi}{11} \lambda_{\phi,eff} M^2_{B-L}
\end{align} 
where $\lambda_{\phi,eff}$ is the physical quartic coupling at the breaking scale $M_{B-L}.$
The ratio of the scalar boson mass to the $B-L$ gauge boson mass is given \cite{IOO} by
\begin{align}
\left( \frac{m_\Phi}{m_{Z'}}   \right)^2 \sim \frac{6}{\pi}  \alpha_{B-L}.  
\label{massrelation}
\end{align}
The condition that the $B-L$ gauge coupling does not diverge  up to the Planck scale
requires $\alpha_{B-L} < 0.015$ at $M_{B-L}$. Hence
the scalar boson becomes  lighter than the $B-L$ gauge boson, $m^2_\Phi < 0.03 \ m^2_{Z'}$.
Such a very light scalar boson  is a general prediction of the CW mechanism.
\subsection{Electroweak symmetry breaking}
The EWSB is triggered by the $B-L$ breaking.
In the previous paper \cite{IOO}, we  assumed a small negative value of the mixing $\lambda_{mix}$
of $H$ and $\Phi$. In this paper, we show that it can be generated radiatively
by solving the RGE (\ref{RGElammix}). In solving the RGE, we put a boundary condition
that  the mixing $\lambda_{mix}$ vanishes at $M_{UV}$.
It comes from the flatness condition of  the scalar potential  in the $H$ direction
at the UV cut-off scale. 

RG evolutions of various couplings can be obtained numerically and 
the behavior of the running of the scalar mixing is drawn in figure 2.
From it, we can read that a very small negative mixing is radiatively induced at IR scale.
In order to understand the universality of such a behavior, we give the
following approximate argument. 
Since $|\lambda_{mix}| \ll 1$,
the RGE is approximated  as
\begin{align}
\frac{d\lambda_{mix}}{dt} \sim \frac{3}{4 \pi^2} g_{mix}^2g_{B-L}^2 .
\label{RGElammix-sim}
\end{align}
If there were no gauge mixing between $U(1)_Y$ and $U(1)_{B-L}$  gauge fields,
the scalar mixing term would  never be generated radiatively.

Let us  look at the RGE of the gauge mixing (\ref{RGEgmix}). 
Since the gauge mixing term is much smaller than other gauge couplings, eq.(\ref{RGEgmix}) is
approximated as
\begin{align}
\frac{d g_{mix}}{dt} \sim \frac{2}{3 \pi^2} g_{B-L} g_Y^2.
\label{RGEgm-sim}
\end{align}
The $\beta$-function is proportional to the cube of the $B-L$ and $U(1)_Y$ gauge couplings.
Hence even if the gauge mixing is absent at some scale, it is radiatively generated.

\begin{figure}[t]\begin{center}
\includegraphics[scale=.6]{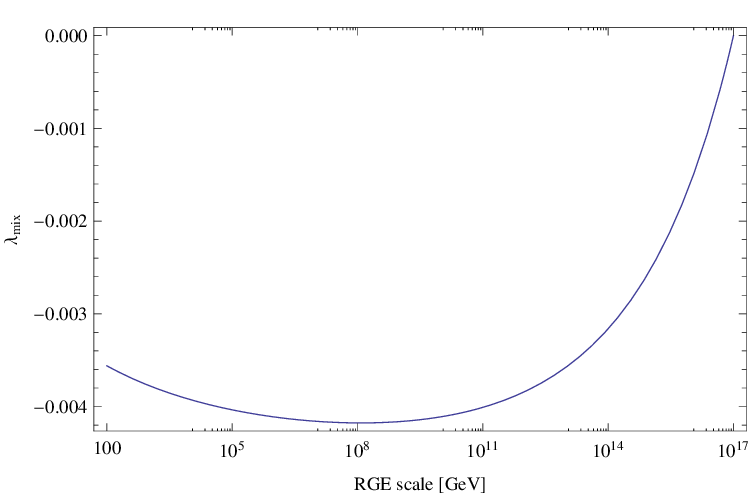}
\caption{
RG evolution of scalar mixing between a SM singlet $\Phi$ and the Higgs $H$.
Starting from zero mixing at $M_{UV}=10^{17}GeV$, a small negative mixing is radiatively
generated at a lower energy scale. The mixing triggers the EWSB.
}
\end{center}
\end{figure}

Now from eq.(\ref{RGElammix-sim}), the scalar mixing $\lambda_{mix}$ is also radiatively generated
through the gauge mixing.
The running of the scalar mixing coupling is drawn in fig. 2.
Because of the very small gauge mixing $g_{mix}$, the scalar mixing is much highly suppressed.
The magnitude of the scalar mixing at a lower energy scale is roughly estimated from (\ref{RGElammix-sim})
and (\ref{RGEgm-sim}) as
\begin{align}
\lambda_{mix} \propto - g_{mix}^2  g_{B-L}^2 \propto -g_{B-L}^4 g_Y^4.
\end{align}
The sign of the scalar mixing is negative at a  lower energy scale because of the positive $\beta$ function in
(\ref{RGElammix-sim}). \footnote{Note that the above argument is valid only when the RGE (\ref{RGElammix}) can be 
approximated by (\ref{RGElammix-sim}). The figure 2 shows that the beta function changes its sign to negative
at a lower energy scale. In this region, the top Yukawa coupling $Y_t$ becomes larger there
and the term proportional to $\lambda_{mix} Y_t^2 (<0)$ becomes dominant. }

If the $\Phi$ field acquires a VEV $\langle \Phi \rangle =M_{B-L}$, the mixing term $\lambda_{mix} (H^\dagger H)(\Phi^\dagger \Phi)$
gives an effective mass term of the $H$ field.
Since  the coefficient $\lambda_{mix}$ is negative, the EWSB is triggered and the Higgs VEV is given by 
\begin{align}
v= \langle H \rangle = \sqrt{\frac{-\lambda_{mix}}{\lambda_H}}  M_{B-L} \sim 
c \frac{\alpha_{B-L} \alpha_Y}{\sqrt{\lambda_H}} M_{B-L}.
\label{Higgsvev}
\end{align}
The coefficient is dependent on the details of the running but roughly given by 
$c \sim 250$. 
This gives the ratio between the EWSB scale to the $B-L$ symmetry breaking scale.
\subsection{Model predictions}
The model has three additional parameters ($g_{B-L}$, $\lambda_{\Phi}$ and $g_{mix}$)
besides the Yukawa couplings of the right-handed neutrinos.
The Yukawa couplings do no affect the dynamics of the scalar potential very much
if they are within the perturbative regime.
Also the gauge mixing $g_{mix}$ is experimentally constrained to be small at a low energy scale
and its magnitude is almost determined by other gauge couplings.
The other two parameters determine the $B-L$ breaking scale $M_{B-L}$ and
EWSB scale $M_{EW}$. The quartic coupling $\lambda_\Phi$ is directly related to $M_{B-L}$
through the RGE of $\lambda_\Phi$.
The other coupling $g_{B-L}$ determines the Higgs VEV through eq.(\ref{Higgsvev}).
So two parameters $\lambda_{\Phi}, g_{B-L}$ determine the two breaking scales
$M_{B-L}$ and $M_{EW}=v$. 

\begin{figure}[t]
\label{figure3}
\begin{center}
\includegraphics[scale=.8]{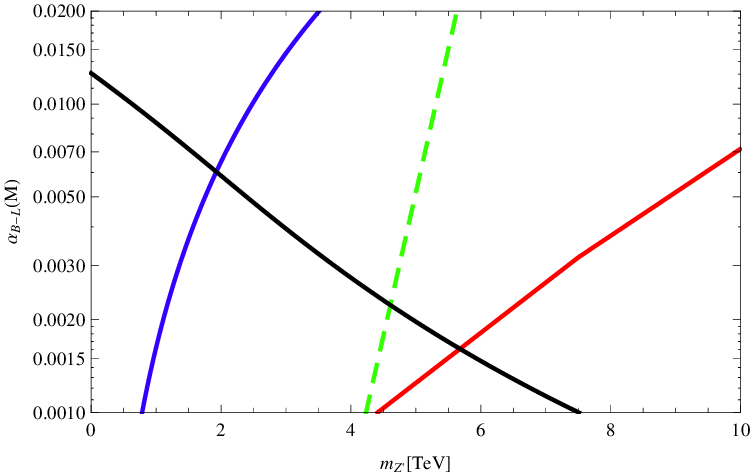}
\caption{Model prediction is drawn in the black line (from top left to down right).
The $B-L$ gauge coupling $\alpha_{B-L}$ and  the  gauge boson mass $m_{Z'}$
are related because of the {\it flat potential assumption at the Planck scale}. 
The left side of the most left solid line in blue has been already excluded by 
 the LEP experiment.
The left of the dashed line can be explored in the 5-$\sigma$ significance 
 at the LHC with $\sqrt{s}$=14 TeV and an integrated 
 luminosity 100 fb$^{-1}$.
The left of the most right solid line (in red) can be explored at the ILC with 
 $\sqrt{s}$=1 TeV, assuming 1\% accuracy. 
}
\end{center}
\end{figure}

The experimental input $v=246$ GeV gives a relation between the two parameters
and the dynamics of the model is essentially described by a single parameter.
The figure 3 shows the prediction of our model.
The vertical axis is the strength of $\alpha_{B-L}$ and the horizontal axis is the
mass of the $B-L$ gauge boson.
The black line (from top left to down right) shows the prediction of our model. 
The model prediction can be heuristically understood as follows.
If we fix the $U(1)_Y$ coupling $g_Y$ and the Higgs quartic coupling $\lambda_H$, 
the input $v=246$ GeV and the relation
(\ref{Higgsvev}) relates the $B-L$ breaking scale $M_{B-L} $ and the gauge coupling 
$g_{B-L}$ as 
$M_{B-L} \sim v \sqrt{\lambda_H}/ (c \alpha_{B-L} \alpha_Y)$.
On the other hand, the gauge boson mass is given by $m^2_{Z'} = 16 \pi \alpha_{B-L} M_{B-L}^2$.
Hence, we have a  relation between the $B-L$ gauge boson mass and the gauge coupling
\begin{align}
\alpha_{B-L} m_{Z'}^2   \sim (16 \pi \lambda_H v^2/c^2 \alpha_{Y}^2) .
\end{align}
The quantities in the rhs are all known.
This gives a very crude approximation of 
 the model prediction of the TeV scale $B-L$ model with a flat potential at the UV scale $M_{UV}$.
A more precise relation is obtained, as shown in the figure 3, by solving the RGEs.
We have also drawn the excluded region by LEP2 experiment,
the LHC reach at 14 TeV with 100 $fb^{-1}$, and the ILC reach at 1 TeV.
\subsection{Higgs quartic coupling and the stability bound}
Finally we consider the RGE of the Higgs quartic coupling. It is assumed to be zero at the UV cut-off scale.
The RGE of the Higgs quartic coupling is given by eq.(\ref{RGElamh}).
Compared to the SM,
the $\beta$ function has a contribution from 
the scalar mixing term $\lambda_{mix}$ and 
the $U(1)$ gauge mixing $g_{mix}$. 
Both mixings contribute to the $\beta$ function  positively. 
\begin{figure}[t]\begin{center}
\includegraphics[scale=.6]{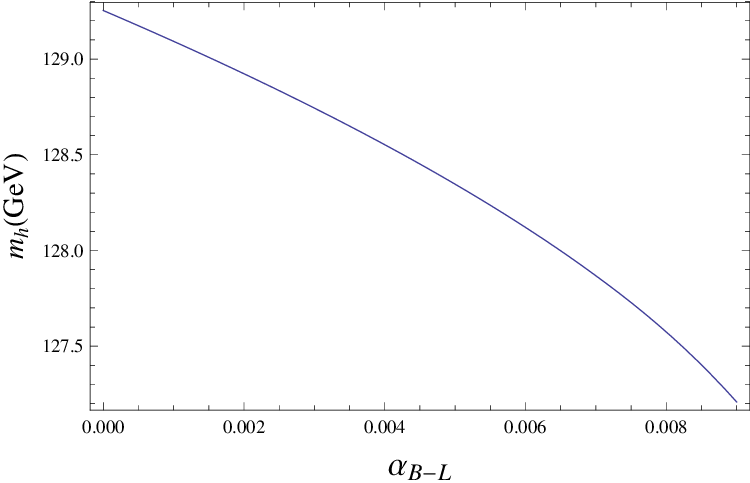}
\caption{
Higgs stability bound up to $M_{UV}=10^{17}$ GeV as a function of the $B-L$ gauge coupling.
We set $M_t=173.1$ GeV and $\alpha_s(M_Z)=0.1184.$ 
}
\end{center}
\end{figure}
Since the $\beta$ function is negative in the SM for the $126$ GeV Higgs mass,
the effect of the $B-L$ gauging reduces the magnitude of the slope of the running coupling.
Hence, the $B-L$ extension of the SM makes the vacuum stabler up to a higher energy scale
than the SM.
Namely, the stability bound (\ref{stability-bound}) of the Higgs boson mass is lowered.

The stability bound is estimated within the 2-loop SM RGEs and the 1-loop $B-L$ RGEs 
including the threshold effects.
The figure 4 shows the Higgs boson mass at the stability bound up to the UV cut off scale 
$M_{UV}=10^{17}$ GeV
as a function of the $B-L$ gauge coupling.
For a stronger $B-L$ gauge coupling, the effect to the $\beta$ function become larger
and the Higgs boson mass at the stability bound can be lowered.
In the present model with a flat Higgs potential at $M_{UV}$, the gauge coupling
and the $B-L$ gauge boson mass are related as shown in figure 3. So $\alpha_{B-L}$ 
must be smaller than $0.006$ and the stability bound can be lowered maximally by about 1 GeV
from the SM prediction. 
\section{Conclusions}
The work is motivated by the
 recent discovery of 126 GeV Higgs-like particle and also non-discovery of the low energy supersymmetric 
particles. The LHC experiment as well as other precision experiments such as the B-factories have put
stringent constraints on the physics beyond the SM. In particular, a large parameter region 
of the low energy supersymmetry has been already excluded.

In this paper, we take an alternative approach to the hierarchy problem, namely, instead of introducing
a large set of particles like in the supersymmetric models, we follow the Bardeen's 
argument on the hierarchy problem and construct a model with the classical conformality.
It  connects the electroweak physics with the Planck scale physics.
A minimal construction of such a model with phenomenological viability is 
the $B-L$ extension of the SM at the TeV scale.
The model has a classical conformality, and  scalar mass terms are absent. 
Therefore, the symmetries must be broken radiatively via the Coleman-Weinberg mechanism.

The TeV scale $B-L$ model is the Occam's razor scenario to solve the hierarchy 
problem, the vacuum stability condition as well as the phenomenological viabilities.
There are two reasons why the $B-L$ extension is necessary.
The first reason is the dynamics of the symmetry breaking.
Since the CW mechanism does not work within the SM, we need 
another sector to achieve the radiative symmetry breaking. 
The $B-L$ gauge interaction is minimal for this purpose. 
Another reason is  the phenomenology.
It is also a minimal extension to explain the neutrino oscillations as well as the leptogenesis.
The breaking scale of the $B-L$ sector in this model 
is required to be not much higher than the TeV scale in order to 
avoid  large logarithmic corrections to the Higgs mass.
In  previous papers \cite{IOO3}, we showed that the TeV scale $B-L$ breaking is compatible with 
the leptogenesis scenario if the masses of the right-handed neutrinos are almost degenerate
and the resonant leptogenesis \cite{resonantLG} can work.

The main analysis of the paper is the RGEs in section 5.
Motivated by the 126 GeV Higgs-like particle, we assume that the Higgs has a flat
potential at the UV scale (e.g., the Planck scale).
We showed that a small negative value of the scalar mixing term between the Higgs
$H$ and the SM singlet scalar $\Phi$ is radiatively generated.
Once the $B-L$ symmetry is spontaneously broken by the CW mechanism at the TeV scale,
the radiatively generated mixing triggers the EWSB.
The ratio between the two breaking scales is dynamically determined in terms of the 
gauge couplings. This gives another reason why the TeV scale is required from the 
EWSB. In most TeV scale $B-L$ models, the scale $M_{B-L}$ of the $B-L$ symmetry
breaking is just assumed by hand. In our model, however, it is  determined
from the relation (\ref{Higgsvev}) as
\begin{align}
M_{B-L} \sim \frac{\sqrt{\lambda_H}}{c \alpha_{B-L} \alpha_Y} \times 246 \ \mbox{GeV}
 \sim \frac{1}{ \alpha_{B-L}} \times 35 \ \mbox{GeV}.
\end{align}
and the mass of the $B-L$ gauge boson is given by
\begin{align}
m_{Z'}= \sqrt{16 \pi \alpha_{B-L}} \ M_{B-L} \sim \frac{1}{\sqrt{\alpha_{B-L}}} \times 250 \ \mbox{GeV}.
\end{align}

The dynamics of the model is essentially controlled by a single parameter, and 
it has a high predictability.
If an extra $U(1)$ gauge boson and a SM singlet scalar are  found in the future,
the prediction of our model is the mass relation (\ref{massrelation}), e.g.,
\begin{align}
m_\phi \sim 0.1 \ m_{Z'}
\end{align}
for $\alpha_{B-L} \sim 0.005$.
The CW mechanism in the $B-L$ sector predicts  a lighter 
 SM singlet Higgs boson than the extra $U(1)$ gauge boson.
It is different from the ordinary TeV scale $B-L$ model where the symmetry 
is broken by a negative squared mass term.

\section*{Acknowledgments} 
The analysis of the paper was first presented by SI at the workshop KEK-TH-2012  on 5 March 2012.
An interested reader can look at the slide, \\ 
\ \ http://research.kek.jp/group/riron/workshop/theory2012/talkfiles/iso.pdf
\\
We thank the participants in the workshop for useful discussions.
We also acknowledge Nobuchika Okada for collaborations in the early stage.
The research is supported in part by Grant-in-Aid for Scientific 
Research (19540316) from MEXT, Japan (S.I), 
Grant-in-Aid for Scientific research on Innovative Areas, No. 23104009 (Y.O).
We are also supported in part by "The Center for the Promotion of Integrated Sciences (CPIS) "  of Sokendai.

\section*{Appendix A}
We list the 1-loop  renormalization group equations of various coupling constants in the $B-L$ model
 \cite{B-LRGE}.
In the body of the paper, we also use  $\alpha_{\lambda,a}=\lambda_a/4 \pi$ for the scalar quartic couplings,
$\alpha_{A}=g_{A}^2/4\pi$ for the gauge couplings and $\alpha_N^i=(Y_N^i)^2/4\pi$
for the Yukawa couplings. \\
\\
\noindent
The RGEs of the gauge couplings are given by
\begin{equation}
 \frac{dg_Y}{dt}=\frac{1}{16\pi^2}\left[\frac{41}{6} g_Y^3\right],
\label{RGEgy}
\end{equation}
\begin{equation}
 \frac{dg_{B-L}}{dt}=\frac{1}{16\pi^2}\left[12g_{B-L}^3
  +2\frac{16}{3}g_{B-L}^2g_{mix}+\frac{41}{6} g_{B-L}g_{mix}^2\right],
\label{RGEgbl}
\end{equation}

\begin{equation}
 \frac{dg_{mix}}{dt}=\frac{1}{16\pi^2}\left[
  \frac{41}{6}g_{mix}\left(g_{mix}^2+2g_Y^2\right)
  +2\frac{16}{3}g_{B-L}\left(g_{mix}^2+g_Y^2\right)
  +12g_{B-L}^2g_{mix}\right].
\label{RGEgmix}
\end{equation}
The RGEs for the Yukawa couplings are
\begin{equation}
 \frac{dY_t}{dt}=\frac{1}{16\pi^2}Y_t\left(
  \frac{9}{2}Y_t^2-8g_3^2-\frac{9}{4}g^2-\frac{17}{12}g_Y^2
  -\frac{17}{12}g_{mix}^2-\frac{2}{3}g_{B-L}^2
  -\frac{5}{3}g_{mix}g_{B-L}\right),
\label{RGEyt}
\end{equation}

\begin{equation}
 \frac{dY_N}{dt}=\frac{1}{16\pi^2}Y_N\left(
  Y_N^2+\frac{1}{2}Tr\left[Y_N^2\right]-6g_{B-L}^2
  \right).
\label{RGEyn}
\end{equation}
Finally the RGEs for the scalar quartic couplings are given by
\begin{eqnarray}
 \frac{d\lambda_H}{dt}=\frac{1}{16\pi^2}\left(
  24\lambda_H^2+\lambda_{mix}^2-6Y_t^4+\frac{9}{8}g^4+\frac{3}{8}g_Y^4
  +\frac{3}{4}g^2g_Y^2+\frac{3}{4}g^2g_{mix}^2
  +\frac{3}{4}g_Y^2g_{mix}^2+\frac{3}{8}g_{mix}^4
\nonumber\right.\\\left.
  +\lambda_H\left(12Y_t^2-9g^2-3g_Y^2-3g_{mix}^2\right)\right),
\label{RGElamh}
\end{eqnarray}

\begin{equation}
 \frac{d\lambda_\Phi}{dt}=\frac{1}{16\pi^2}\left(
  20\lambda_\Phi^2+2\lambda_{mix}^2-\frac{1}{2}Tr\left[Y_N^4\right]
  +96g_{B-L}^4+\lambda_\Phi\left(2Tr\left[Y_N^2\right]
  -48g_{B-L}^2\right)\right),
\label{RGElamphi}
\end{equation}

\begin{eqnarray}
 \frac{d\lambda_{mix}}{dt}=\frac{1}{16\pi^2}\left[\lambda_{mix}\left(
  12\lambda_H+8\lambda_\Phi+4\lambda_{mix}+6Y_t^2-\frac{9}{2}g^2
  -\frac{3}{2}g_Y^2-\frac{3}{2}g_{mix}^2
  +Tr\left[Y_N^2\right]-24g_{B-L}^2\right)
\nonumber\right.\\\left.
  +12g_{mix}^2g_{B-L}^2\right].
\label{RGElammix}
\end{eqnarray}



\end{document}